\documentclass[11pt,a4paper,english,american]{article}
\pdfoutput=1
\usepackage{lmodern}

\usepackage[T1]{fontenc}
\usepackage[latin9]{inputenc}
\usepackage{amsmath}
\usepackage{amssymb}
\usepackage{esint}

\RequirePackage[colorlinks=true
,urlcolor=blue
,anchorcolor=blue
,citecolor=blue
,filecolor=blue
,linkcolor=blue
,menucolor=blue
,linktocpage=true
,pdfproducer=medialab
]{hyperref}

\makeatletter

\usepackage{jcappub}
\numberwithin{equation}{section}

\pdfoutput=1
\renewcommand\[{\begin{equation}}
\renewcommand\]{\end{equation}}

\AtBeginDocument{
  
}

\makeatother

\usepackage{babel}

\begin{document}

\title{Imperfect Dark Matter }

\author{Leila Mirzagholi}

\author{and Alexander Vikman}

\affiliation{Arnold Sommerfeld Center for Theoretical Physics, \\
Ludwig Maximilian University Munich, \\
Theresienstr. 37, D-80333, Munich, Germany}

\emailAdd{l.mirzagholi@physik.uni-muenchen.de, alexander.vikman@lmu.de}

\abstract{We consider cosmology of the recently introduced \emph{mimetic matter
}with higher derivatives\emph{ }(HD)\emph{.} Without HD this system
describes irrotational dust \textendash{} Dark Matter (DM) as we see
it on cosmologically large scales. DM particles correspond to the
shift-charges \textendash{} Noether charges of the shifts in the field
space. Higher derivative corrections usually describe a deviation
from the thermodynamical equilibrium in the relativistic hydrodynamics.
Thus we show that\emph{ mimetic matter }with HD corresponds to an
\emph{imperfect} DM which: i) renormalises the Newton's constant in
the Friedmann equations, ii)\emph{ }has zero pressure when there is
no extra matter in the universe, iii) survives the inflationary expansion
which puts the system on a dynamical attractor with a vanishing shift-charge,
iv) perfectly \emph{tracks} \emph{any }external matter on this attractor,
v) can become the main (and possibly the only) source of DM, provided
the shift-symmetry in the HD terms is broken during some small time
interval in the radiation domination époque. \\
In the second part of the paper we present a hydrodynamical description
of general anisotropic and inhomogeneous configurations of the system.
This imperfect \emph{mimetic fluid} has an energy flow in the field's
rest frame. We find that in the Eckart and in the Landau-Lifshitz
frames the \emph{mimetic fluid} possesses nonvanishing vorticity appearing
already at the first order in the HD. Thus, the structure formation
and gravitational collapse should proceed in a rather different fashion
from the simple irrotational DM models. }

\subheader{LMU-ASC 08/15}

\maketitle

\section{Introduction}

The origin of Dark Matter (DM) is one of the oldest and biggest puzzles
in cosmology and particle physics. Surprisingly, even \emph{basic
macroscopic} nature of DM is not yet understood \textendash{} indeed,
we do not know whether DM is a gas of some particles%
\footnote{This gas can be composed of the so-called Weakly Interacting Massive
Particles (WIMPs) \cite{Steigman:1984ac}, sterile right-handed neutrinos
\cite{Dodelson:1993je,Asaka:2005an,Boyarsky:2012rt} and even different,
and some times numerous copies of the SM, see e.g. \cite{Dvali:2009fw,Foot:2014mia}
just to mention few options. Whether the axion DM \cite{Abbott:1982af,Dine:1982ah,Preskill:1982cy}
represents a condensate or not is still debated, for the most recent
discussion see \cite{Guth:2014hsa}. %
} beyond the Standard Model (SM), or a gas of primordial black holes
see e.g. \cite{Carr:1974nx,Hawkins:2011qz}, or some other \emph{macroscopic}
objects, Q-Balls \cite{Coleman:1985ki}, Topological defects etc,
see e.g. \cite{Jacobs:2014yca,Kusenko:1997si,Kusenko:2001vu,Murayama:2009nj,Derevianko:2013oaa,Stadnik:2014cea},
or some fluid e.g. \cite{Peebles:2000yy} or Bose-Einstein condensates
\textendash{} some classical scalar fields e.g. \cite{Hu:2000ke},
\cite{ArmendarizPicon:2005nz,Khoury:2014tka} or even some effective
solid \cite{Bucher:1998mh}. In the latter approach, where one assumes
high occupation numbers of some new fields, we can also incorporate
the relativistic version of MOND \cite{Milgrom:1983ca,Famaey:2011kh,Bruneton:2007si}
- TeVeS \cite{Bekenstein:2004ne} and numerous other modifications
of general relativity (GR), e.g. \cite{Zlosnik:2006zu}. The simplest
modification of GR can be achieved by promoting it to a scalar-tensor
theory. 

GR enjoys a very powerful symmetry \textendash{} diffeomorphism invariance.
One of the manifestations of its power is that one can \emph{parametrize}
the metric $g_{\mu\nu}$ by a scalar field $\varphi$ and an auxiliary
metric $\ell_{\mu\nu}$ in a general \emph{disformal} way \cite{Bekenstein:1992pj}
\begin{equation}
g_{\mu\nu}=C\left(\varphi,X\right)\ell_{\mu\nu}+D\left(\varphi,X\right)\varphi_{,\mu}\varphi_{,\nu}\,,\label{eq:Disformal}
\end{equation}
where $X=\tfrac{1}{2}\ell^{\mu\nu}\varphi_{,\mu}\varphi_{,\nu}$ and
$C\left(\varphi,X\right)$ and $D\left(\varphi,X\right)$ are free
functions%
\footnote{There is a well known physical disformal transformation\textendash{}the
effective / acoustic metric for the propagation of small perturbations
in k-\emph{essence }\cite{ArmendarizPicon:1999rj,ArmendarizPicon:2000ah,ArmendarizPicon:2000dh}
or irrotational hydrodynamics is a particular disformal transformation
of the gravitational metric $g_{\mu\nu}$, see \cite{Moncrief:1980,Babichev:2007dw}. %
}, and obtain the Einstein equations (for $g_{\mu\nu}$) by variation
of the action with respect to $\varphi$ and $\ell_{\mu\nu}$ instead
of $g_{\mu\nu}$, see \cite{Deruelle:2014zza}. The only exception
from this rule corresponds to a singular parameterisation when \cite{Deruelle:2014zza}
\[
D\left(\varphi,X\right)=f\left(\varphi\right)-\frac{C\left(\varphi,X\right)}{2X}\,.
\]
When the transformation is singular, there are new degrees of freedom
and new physics modifying GR. \emph{Mimetic Dark Matter} \cite{Chamseddine:2013kea}
is one of the theories of this type and makes use of the transformation
\eqref{eq:Disformal} with $C=2X$ and $D=0$, so that 
\begin{equation}
g_{\mu\nu}=\left(\ell^{\alpha\beta}\varphi_{,\alpha}\varphi_{,\beta}\right)\ell_{\mu\nu}\,.\label{eq:Mimetic_Asatz}
\end{equation}
 It is important that in this case the system is Weyl invariant with
respect to the transformations of the auxiliary metric $\ell_{\mu\nu}$.
Soon it was realised that \emph{Mimetic Dark Matter} is equivalent
to the fluid description of irrotational dust \cite{Golovnev:2013jxa,Barvinsky:2013mea}
with the \emph{mimetic field} $\varphi$ playing the role of the velocity
potential. Models of this type also appear in the IR limit of the
projectable version of Ho\v{r}ava-Lifshitz gravity \cite{Horava:2009uw,Mukohyama:2009mz,Blas:2009yd}
and correspond to a scalar version of the so-called Einstein Aether
\cite{Jacobson:2000xp}. Surprisingly these models can also emerge
in the non-commutative geometry \cite{Chamseddine:2014nxa}. In \cite{Chamseddine:2014vna}
this class of systems was further extended by i) adding a potential
$V\left(\varphi\right)$ which allows to obtain an \emph{arbitrary
}equation of state for this dust-like matter with zero sound speed,
as it was done earlier in \cite{Lim:2010yk}; ii) by introducing higher
derivatives (HD) which provide a nonvanishing sound speed. The latter
modification allowed one to study inflationary models with the creation
of quantum cosmological perturbations. Moreover, this finite sound
speed can suppress the structure on small scales \cite{Capela:2014xta}
and have other interesting phenomenological consequences. 

This paper is organised as follows: first, in section \ref{sec:Main-setup}
we present the main equations and reformulate the model from \cite{Chamseddine:2014vna}
in terms of two scalar fields and also we extend this setup by introducing
shift-symmetry breaking in HD terms. In the next section \ref{sec:Cosmology}
we consider background cosmology. In particular, we present constraints
on the models parameters. Our main point in this section is a new
mechanism which allows \emph{mimetic fluid }to survive inflation and
later becomes the main source of DM. This mechanism is based on the
shift-symmetry breaking in the HD terms operating during some cosmologically
short period of time within the radiation-domination époque or possibly
as early as within reheating. Then in section \ref{sec:Fluid-picture}
we consider fluid-like description for the system in the case of shift-symmetry.
The most important point there is that the \emph{mimetic fluid} does
have vorticity and moreover is not a perfect fluid. Indeed the four-velocity
of particles (the Eckart frame) does not coincide with the four-velocity
of energy (Landau-Lifshitz frame). This is very important for studies
of DM on nonlinear scales. Then in section \ref{sec:Conclusions-and-Open}
we discuss our results and mention open question which provide directions
for future research.

\section{Main setup\label{sec:Main-setup} }

The matter part of the action 
\begin{equation}
S\left[g,\lambda,\varphi\right]=\int\mbox{d}^{4}x\,\sqrt{-g}\left(\lambda\left(g^{\mu\nu}\partial_{\mu}\varphi\partial_{\nu}\varphi-1\right)+\frac{1}{2}\gamma\left(\varphi\right)\left(\Box\varphi\right)^{2}\right)\,,\label{eq:action}
\end{equation}
where $\lambda$ is a Lagrange multiplier field, $\varphi$ is the
\emph{mimetic} field, $\gamma\left(\varphi\right)$ is a function
of the field and $\Box=g^{\mu\nu}\nabla_{\mu}\nabla_{\nu}$ with $\nabla_{\mu}\left(\,\,\right)=\left(\,\,\right)_{;\mu}$
being the covariant derivative. This action with a constant $\gamma$
was introduced in \cite{Chamseddine:2014vna} and it generalises \cite{Lim:2010yk,Chamseddine:2013kea},
see also \cite{Haghani:2014ita}. Without higher derivatives this
action describes irrotational dust.%
\footnote{For the Lagrangian description of dust with vorticity see, \cite{Brown:1992kc,Brown:1994py}%
} . We will assume the standard minimal coupling to gravity. Also we
will assume that matter is not directly coupled%
\footnote{This condition can be easily violated, if one applies the \emph{mimetic}
ansatz \eqref{eq:Mimetic_Asatz} in $f\left(R\right)$ theories \cite{Nojiri:2014zqa}. %
} to the \emph{mimetic} field $\varphi$. Throughout the paper we will
either use the units where $8\pi G_{N}=1$ or write $G_{N}$ explicitly
where it is useful for discussion. We also use the signature convention
$\left(+,-,-,-\right)$. Here we have omitted the boundary terms which
are needed because of the higher derivative structure of the theory. 

The Lagrange multiplier $\lambda$ enforces the constraint 
\begin{equation}
g^{\mu\nu}\partial_{\mu}\varphi\partial_{\nu}\varphi=1\,.\label{eq:constaraint}
\end{equation}
Similarly to \cite{Capela:2014xta} and \cite{Haghani:2014ita} one
could also add a term $\nabla_{\mu}\nabla_{\nu}\phi\nabla^{\mu}\nabla^{\nu}\phi$,
to the Lagrangian. However, this term only introduces a direct non-minimal
coupling to gravity 
\[
\int\mbox{d}^{4}x\,\sqrt{-g}\left(\nabla_{\mu}\nabla_{\nu}\varphi\right)\left(\nabla^{\mu}\nabla^{\nu}\varphi\right)=\int\mbox{d}^{4}x\,\sqrt{-g}\left(\left(\Box\varphi\right)^{2}-R^{\mu\nu}\nabla_{\mu}\varphi\,\nabla_{\nu}\varphi\right)\,,
\]
where we have again omitted the boundary terms%
\footnote{By a direct calculation on can find that the non-minimal term $R^{\mu\nu}\partial_{\mu}\varphi\partial_{\nu}\varphi$
induces the anisotropic stress and changes the speed of propagation
of gravitons in cosmology, see e.g. \cite{Saltas:2014dha}. Also the
energy momentum tensor would directly include the Riemann tensor and
another term mimicking the shear-viscosity. %
}. We will not consider this term in this paper. One can use an equivalent
action to describe the dynamics of the system without use of the higher
derivatives 
\[
S'\left[g,\lambda,\varphi,\theta\right]=\int\mbox{d}^{4}x\,\sqrt{-g}\left[\lambda\left(\varphi^{,\mu}\varphi_{,\mu}-1\right)-\gamma\left(\varphi\right)\left(\varphi_{,\mu}\theta^{,\mu}+\frac{1}{2}\theta^{2}\right)-\gamma'\left(\varphi\right)\theta\,\varphi^{,\mu}\varphi_{,\mu}\right]\,,
\]
where $\theta$ is an auxiliary field. The equation of motion for
this field gives $\theta=\Box\varphi$. In addition, there is an equation
of motion for $\varphi$ which can be conveniently written in form
of the current (non) conservation 
\begin{equation}
\nabla_{\mu}J^{\mu}=\frac{1}{2}\gamma'\left(\varphi\right)\theta^{2}\,,\label{eq:Phi_EOM}
\end{equation}
where the the prime denotes the derivative $\partial/\partial\varphi$
and the current is given by 
\begin{equation}
J_{\mu}=\left(2\lambda-\gamma'\left(\varphi\right)\theta\right)\partial_{\mu}\varphi-\gamma\partial_{\mu}\theta\,.\label{eq:Noether_General}
\end{equation}
When the theory is shift-invariant: symmetric with respect to $\varphi\rightarrow\varphi+c$
the current is conserved and is the Noether current corresponding
to the shift symmetry. 

The energy-momentum tensor (EMT) is
\begin{align}
 & T_{\mu\nu}=\frac{2}{\sqrt{-g}}\frac{\delta S}{\delta g^{\mu\nu}}=2\left(\lambda-\gamma'\left(\varphi\right)\theta\right)\,\varphi_{,\mu}\varphi_{,\nu}+\label{eq:EMT_General}\\
 & +\gamma\left(\varphi\right)\left[g_{\mu\nu}\left(\varphi_{,\alpha}\theta^{,\alpha}+\frac{1}{2}\theta^{2}+\frac{\gamma'\left(\varphi\right)}{\gamma\left(\varphi\right)}\theta\right)-\varphi_{,\mu}\theta_{,\nu}-\varphi_{,\nu}\theta_{,\mu}\right]\,,\nonumber 
\end{align}
 where we have assumed that the constraint \eqref{eq:constaraint}
is satisfied together with the equation of motion 
\begin{equation}
\theta=\Box\varphi=\frac{1}{\sqrt{-g}}\partial_{\mu}\left(\sqrt{-g}\, g^{\mu\nu}\partial_{\nu}\varphi\right)\,.\label{eq:teta}
\end{equation}

\section{Cosmology\label{sec:Cosmology}}

Let us consider a spatially-flat Friedmann universe. From the constraint
\eqref{eq:constaraint} it follows that $\varphi=t+c$. Hence the
auxiliary field \eqref{eq:teta} reads 

\begin{equation}
\theta=3H\,,\label{eq:teta_cosmology}
\end{equation}
where $H$ is the Hubble parameter. From the EMT \eqref{eq:EMT_General}
we obtain the energy density 
\begin{equation}
\varepsilon=T_{0}^{0}=2\lambda-\dot{\gamma}\theta-\gamma\left(\dot{\theta}-\frac{1}{2}\theta^{2}\right)\,,\label{eq:energy_density_cosmology}
\end{equation}
and the pressure 
\begin{equation}
p=-\frac{1}{3}T_{i}^{i}=-\dot{\gamma}\theta-\gamma\left(\dot{\theta}+\frac{1}{2}\theta^{2}\right)\,.\label{eq:pressure_cosmology}
\end{equation}
Thus the enthalpy density is 
\[
h=\varepsilon+p=2\lambda-\dot{\gamma}\theta-2\gamma\dot{\theta}=2\lambda-3\dot{\gamma}H-6\gamma\dot{H}\,.
\]
The second term in the last equality is similar to the so-called \emph{inhomogeneous
equation of state} considered in \cite{Nojiri:2005sr,Bamba:2012cp}.
While the charge density of the current \eqref{eq:Noether_General}
is given by 
\begin{equation}
n=J^{0}=2\lambda-\dot{\gamma}\theta-\gamma\dot{\theta}\,.\label{eq:charge_cosmology}
\end{equation}
For the $\varphi$ equation of motion \eqref{eq:Phi_EOM} or current
(non)-conservation we can write 
\begin{equation}
\frac{d}{dt}\left(na^{3}\right)=\frac{9}{2}a^{3}\dot{\gamma}H^{2}\,.\label{eq:charge generation}
\end{equation}
For the Hubble parameter we have the Friedmann equation 
\begin{equation}
H^{2}=\frac{1}{3}\left(\varepsilon+\rho_{\text{ext}}\right)\,,\label{eq:Fredmann1}
\end{equation}
\begin{equation}
\dot{H}=-\frac{1}{2}\left(\varepsilon+\rho_{\text{ext}}+p+P_{\text{ext}}\right)\,,\label{eq:Freidmann2}
\end{equation}
where $\rho_{\text{ext}}$ , $P_{\text{ext}}$ is the external energy
density and pressure of e.g. radiation etc. 

Let us for a while neglect the explicit dependence of $\gamma\left(\varphi\right)$
and consider the case with the shift-symmetry. We will come back to
the breaking of the shift symmetry in the subsection \ref{sub:Transition}.
In that case it was shown in \cite{Chamseddine:2014vna} that the
density perturbations inquire a sound speed 
\begin{equation}
c_{\text{S}}^{2}=\frac{\gamma}{2-3\gamma}\,.\label{eq:sound speed}
\end{equation}
Using the Friedmann equations \eqref{eq:Fredmann1}, \eqref{eq:Freidmann2}
in the shift-symmetric case we obtain that 
\[
\varepsilon=2\lambda+\frac{3}{2}\gamma\left(2\varepsilon+2\rho_{\text{ext}}+p+P_{\text{ext}}\right)\,,
\]
and the pressure is 
\[
p=\gamma\frac{3}{2}\left(p+P_{\text{ext}}\right)\,,
\]
so that 
\[
p=\frac{3\gamma}{2-3\gamma}\, P_{\text{ext}}=3c_{\text{S}}^{2}P_{\text{ext}}\,.
\]
In particular, if the external pressure is vanishing, the pressure of
the mimetic fluid is vanishing as well.  This implies that
when the mimetic fluid is the only matter in the universe, it will
always behave like dust or DM from the point of view of the background
evolution%
\footnote{Note that the presence of dust-like cosmological solutions was showed
in \cite{Haghani:2014ita}, here we prove that there are no other
solutions and universe filled with the shift-symmetric mimetic fluid
always undergoes the cosmological expansion corresponding to the matter
domination époque. %
}.

\subsection{Attractor and ideal tracking }

In the shift-symmetric case it is convenient to exclude $\lambda$
and write the energy density as 
\begin{equation}
\varepsilon=\frac{2}{2-3\gamma}n+\frac{3\gamma}{2-3\gamma}\,\rho_{\text{ext}}\,.\label{eq:general_energy_density_off_attractor}
\end{equation}
In cosmology the shift-charge density will be redshifted as 
\[
n\propto a^{-3}\,.
\]
Thus the whole cosmological effect on the background level is an addition
of the DM like component $ $$2n/\left(2-3\gamma\right)$ and second
part which ideally tracks the external matter. The latter tracking
is just a rescaling of the Newton's constant 
\begin{equation}
G_{\text{eff}}=G_{N}\left(1+\frac{3\gamma}{2-3\gamma}\right)=G_{N}\left(1+3c_{\text{S}}^{2}\right)\,.\label{eq:rescaling_G}
\end{equation}
Thus instead of $G_{N}$ the cosmological expansion will depend on
$G_{\text{eff}}$. 

However, on the level of perturbations, during the radiation-dominated
époque the mimetic fluid adds a component with the sounds speed $c_{\text{S}}^{2}=\gamma/\left(2-3\gamma\right)$
and with equation of state of radiation $w=1/3$.  Indeed, if the shift-symmetry was not broken during inflation and after that, the charge density will be redshifted as $n\propto a^{-3}$, producing at the end a field configuration without any charge, $n = 0$, with a tremendous exponential precision. In that
case from \eqref{eq:charge_cosmology} we obtain 
\begin{equation}
2\lambda_{\star}=\gamma\dot{\theta}_{\star}=3\gamma\dot{H}\,,\label{eq:no charge}
\end{equation}
while from \eqref{eq:general_energy_density_off_attractor} we see
that surprisingly the energy density is not completely redshifted
and is proportional to the external energy density (of radiation,
additional DM etc): 
\begin{equation}
\varepsilon_{\star}=\frac{3\gamma}{2-3\gamma}\,\rho_{\text{ext}}=3c_{\text{S}}^{2}\rho_{\text{ext}}\,.\label{eq:Energy density on attractor}
\end{equation}
Thus the shift-symmetric mimetic fluid cannot be the only source of
DM. 

Clearly $\gamma\sim10^{-10}$ as it is chosen in \cite{Capela:2014xta}
to suppress the power spectrum on sufficiently small wavelengths would
provide a completely unobservable part of DM, if the mimetic fluid
is always shift-symmetric. 

For the pressure we still have 
\[
p_{\star}=\frac{3\gamma}{2-3\gamma}\, P_{\text{ext}}=3c_{\text{S}}^{2}P_{\text{ext}}\,,
\]
hence the neutral mimetic fluid has exactly the same equation of state
as the external matter%
\footnote{This the opposite situation from the anti-tracking happening in \cite{Pujolas:2011he}.%
} 
\[
\frac{p_{\star}}{\varepsilon_{\star}}=\frac{P_{\text{ext}}}{\rho_{\text{ext}}}\,.
\]
Thus the only background manifestation of the mimetic fluid is the
rescaling of the Newton's constant \eqref{eq:rescaling_G}. However,
the perturbations will be rather different because the sound speed
for the \emph{mimetic fluid} is generically rather different from
e.g. radiation \eqref{eq:sound speed}. The \emph{mimetic fluid} can
mimic radiation provided $\gamma=1/3$ with $c_{\text{S}}^{2}=1/3$,
while $\gamma=1/2$ corresponds to the speed of light and the speed
of sound of a canonical scalar field. Such large $\gamma$ would not
be allowed, because of the bounds on the $\delta G_{\text{N}}$ and
because the system in that case would not cluster like the ordinary
CDM.

\subsection{Bounds on the sound speed and abundance\label{sub:Bounds-on-the} }

During the big bang nucleosynthesis (BBN) the bounds on the $\delta G_{N}$
imply that: 
\[
\left.c_{\text{S}}^{2}\right|_{\text{BBN}}\lesssim0.02\,,
\]
if we follow the results from \cite{Rappaport:2007ct}%
\footnote{An earlier work \cite{Bambi:2005fi} provides a weaker bound $c_{\text{S}}^{2}\lesssim0.1$. %
}. Thus from \cite{Rappaport:2007ct} it follows that shift-symmetric
mimetic fluid can only provide up to $6\%$ of the DM and build the
same $6\%$ of the radiation. 

We should stress that we do not know yet what will be the effective
Newton's constant in the Solar System in the presence of the mimetic
fluid. This would require knowledge of the profile of mimetic fluid
around solar system. This task goes beyond the scope of the current
paper. Thus this bound and the bounds below on $\delta G_{N}$ should
be taken with caution. 

If $\gamma$ changes during the matter / radiation equality, but before
that and some time after that the mimetic fluid is practically uncharged
$n\simeq0$, one can use data from the high resolution CMB and find
that \cite{Narimani:2014zha} 
\begin{equation}
3\left(\left.c_{\text{S}}^{2}\right|_{\text{matter}}-\left.c_{\text{S}}^{2}\right|_{\text{radiation}}\right)\lesssim0.105\pm0.049\,,\label{eq:Bound on C without BAO}
\end{equation}
or if one includes baryonic acoustic oscillations: 
\begin{equation}
3\left(\left.c_{\text{S}}^{2}\right|_{\text{matter}}-\left.c_{\text{S}}^{2}\right|_{\text{radiation}}\right)\lesssim0.066\pm0.039\,.\label{eq:Bound on C_BAO}
\end{equation}

Moreover, the scalar / irrotational DM (IDM) can be interesting for
seeding the primordial black holes (BH), \cite{Sawicki:2013wja}.
In this context it was shown that to accelerate the formation of the
primordial black holes one has to impose $c_{\text{S}}^{2}\lesssim\Phi\simeq10^{-5}$.
In that case essentially all primordial overdensities of IDM collapse
to black holes. Thus the abundance of IDM with respect to CDM was
showed to be $\varepsilon_{\text{IDM}}/\rho_{\text{CDM}}\lesssim10^{-7}$,
because of the ratio between the mass of the Sagittarius A{*} and
the mass of the Milky Way. In particular this bound on abundance would
be applicable for $\gamma\simeq10^{-10}$ taken in \cite{Capela:2014xta},
provided \emph{mimetic fluid} were irrotational and perfect. This
would place a by far more stringent bound 
\begin{equation}
\gamma\simeq c_{\text{S}}^{2}\lesssim3\times10^{-8}\,.\label{eq:bound on gamma}
\end{equation}
Note that in this work we are discussing a model which can survive
the inflationary red-shifting, whereas energy density for the IDM
models studied in \cite{Sawicki:2013wja} would completely disappear
during inflation. 

Further we should mention that, if $c_{\text{S}}^{2}\geq10^{-5}$
during the matter domination era the formation of the structure and
black holes is not that efficient and the bound $\varepsilon_{\text{IDM}}/\rho_{\text{CDM}}\lesssim10^{-7}$
is not applicable. 

On the other hand there are constraints on the sound speed of DM \cite{Sandvik:2002jz}
where it was claimed that the bound is $c_{\text{S}}^{2}<10^{-5}$.
We expect that the nonlinear collapse proceeds radically differently
in the \emph{mimetic imperfect fluids}. Indeed, as we show in the
next section \ref{sec:Fluid-picture}, contrary to the models from
\cite{Sawicki:2013wja}, our \emph{imperfect} DM has an intrinsic
anisotropy in form of the energy flow $q_{\mu}$ \eqref{eq:energy flux}
thus the system is not a perfect fluid. Moreover, there is an intrinsic
vorticity. Thus the bound \eqref{eq:bound on gamma} is not directly
applicable.

\subsection{Transition \label{sub:Transition}}

Now let us consider a short period of time during the radiation domination
époque when the shift-symmetry is broken and the shift-charge can
be generated. Thus the theory is shift-symmetric, before and after
this period of time which is finished at $t_{\text{cr}}$. Suppose,
for simplicity, that the charge is generated during $\Delta t\ll H^{-1}$
so that we can neglect the cosmological evolution during the charge
generation. During the creation of the charge we can assume that the
\emph{mimetic fluid} is completely subdominant in comparison with
the radiation. Then equation for the charge generation \eqref{eq:charge generation}
gives us
\[
na^{3}=\frac{9}{2}\int_{(t_{\text{cr}}-\Delta t)}^{t}dt'\, a^{3}\dot{\gamma}H^{2}\simeq\frac{3}{2}\int_{(t_{\text{cr}}-\Delta t)}^{t}dt'a^{3}\dot{\gamma}\rho_{\text{rad}}\simeq\frac{3}{2}a^{3}\rho_{\text{rad}}\left(t_{\text{cr}}\right)\Delta\gamma\,,
\]
where in the last approximate equality we have used the sudden change
$\Delta t\ll H^{-1}$ approximation. Hence, the created charge density
is 
\[
n\left(t_{\text{cr}}\right)\simeq\frac{3}{2}\rho_{\text{rad}}\left(t_{\text{cr}}\right)\Delta\gamma\,.
\]
During the latter times and, in particular, during the matter domination
$\gamma=\gamma_{\text{early}}+\Delta\gamma\ll1$. For example the
following $\gamma\left(\varphi\right)$ would work 
\[
\gamma\left(\varphi\right)=\gamma_{\text{early}}+\frac{1}{2}\Delta\gamma\left(\tanh\left(\frac{\varphi-\varphi_{\text{cr}}+\Delta\varphi}{\Delta\varphi}\right)+1\right)\,.
\]
After the charge is created and the shift-symmetry is restored, the
energy density in the mimetic fluid redshifts as DM. In particular,
\[
n\left(t\right)=n\left(t_{\text{cr}}\right)\left(\frac{a_{\text{cr}}}{a\left(t\right)}\right)^{3}\,.
\]
If the transition generates the same amount of energy density in the
mimetic fluid as it would be in the standard CDM picture, then $\varepsilon=\rho_{\text{eq}}\left(a/a_{\text{eq}}\right)^{-3}$
where $\rho_{\text{eq}}$ is the energy density in radiation at the
moment of matter-radiation equality and $a_{\text{eq}}$ is the scale
factor at this moment. Using \eqref{eq:general_energy_density_off_attractor}
we obtain that 
\[
\Delta\gamma=\frac{2}{3}\left(\frac{a_{\text{cr}}}{a_{\text{eq}}}\right)\simeq\frac{2}{3}\,\frac{z_{\text{eq}}}{z_{\text{cr}}}\,,
\]
where $a_{\text{cr}}=a\left(t_{\text{cr}}\right)$, $z_{\text{eq}}\sim10^{4}$
is the redshift of the matter-radiation equality, and $z_{\text{cr}}$
corresponds to the redshift at the moment of charge creation. Thus
the earlier happens the creation of charge the smaller is the change
of $\gamma$ and of the the sound speed needed. Note that from the
bounds on the change of the Newton's constant \eqref{eq:Bound on C_BAO}
one obtains that the charge creation should happen at the redshift
$z\gtrsim10^{6}$ from now which corresponds to the temperature $T\gtrsim10^{2}\,\mbox{eV}$.
In particular, for the choice of \cite{Capela:2014xta} $\Delta\gamma\lesssim10^{-10}$
which was interesting for the small scales phenomenology%
\footnote{Note that in normal units this means $\Delta\gamma\lesssim\left(10^{-5}M_{\text{Pl}}\right)^{2}\simeq\left(10^{13}\,\mbox{GeV}\right)^{2}$.%
} (to suppress perturbations with wavelengths below 100 kpc), $z\gtrsim10^{14}$
or temperature $T\gtrsim10\,\mbox{GeV}$. Thus there the charge creation
happens before the quark\textendash{}gluon transition and can easily
occur during the electroweak phase transition. In general the temperature
at the transition is related to $\Delta\gamma$ as 
\[
T_{\text{cr}}\simeq\frac{T_{\text{eq}}}{\Delta\gamma}\simeq\frac{\mbox{eV}}{\Delta\gamma}\,.
\]
Thus to move the temperature to the region of physics currently not
probed by accelerators, $T_{\text{cr}}\gtrsim100\,\mbox{TeV}$, we
have to assume that $\Delta\gamma\lesssim10^{-14}$. In general, it
would be rather interesting to connect this charge-creation moment
with a phase transition in the early universe or with reheating after
inflation. The sounds speed corresponding to the efficient clustering
is $c_{\text{S}}^{2}\lesssim10^{-5}$ \cite{Sandvik:2002jz}, \cite{Sawicki:2013wja}
thus $\Delta\gamma\lesssim\gamma\lesssim10^{-5}$, so that $T_{\text{cr}}\gtrsim0.1\,\mbox{MeV}$
which is between the primordial nucleosynthesis and electron-positron
annihilation. \\

Here it is important to note that a potential term $V\left(\varphi\right)$
would not help to generate the charge without $\gamma\left(\varphi\right)$.
Indeed, if $n\left(t_{i}\right)=0$, then  
\[
n\left(t\right)=\int_{t_{i}}^{t}dt'\dot{V}\left(\frac{a\left(t'\right)}{a\left(t\right)}\right)^{3}<\Delta V\,,
\]
where in the last inequality we have used the fact that the universe
is expanding. Therefore, the increase of the cosmological constant
would be larger than the produced charge density making it impossible
for the mimetic fluid to become the main source of DM. Of course one
could use changes of the cosmological constant (during say phase transitions
provided $\Delta V>0$) to generate some of the DM density. But clearly
it is not enough to explain the hole DM abundance. 

On the other hand, one can invert the argument \textendash{} exactly
during this short phase the breaking of the shift-symmetry can generate
a change in the potential, $\Delta V$, i.e. the observed small cosmological
constant. This creation of $\Lambda$ is accompanied with a production
of the negligible amount of the shift-charge. 

This mechanism of DM production, potentially can introduce some amount
of isocurvature cosmological perturbations. In a forthcoming publication
we will show that generically these isocurvature perturbations are
negligible.

\section{Fluid picture\label{sec:Fluid-picture} }

In this chapter we consider unbroken shift-symmetry: $\varphi\rightarrow\varphi+c$,
so that $\gamma=const$. We will follow the reviews \cite{Andersson:2006nr,Gourgoulhon:2006bn}
and the discussion of a similar system \textendash{} the imperfect
fluid from \emph{Kinetic Gravity Braiding} \cite{Pujolas:2011he}. 

The shift-symmetry Noether current can be decomposed using the local
rest frame (LRF) given by a natural choice $u_{\mu}=\partial_{\mu}\varphi$
as 
\begin{equation}
J_{\mu}=nu_{\mu}-\gamma\bot_{\mu}^{\lambda}\nabla_{\lambda}\theta\,,\label{eq:Noether decomposed}
\end{equation}
where the shift-charge density is 
\begin{equation}
n=2\lambda-\gamma\dot{\theta}\,,\label{eq:Charge density}
\end{equation}
with the notation $u^{\mu}\nabla_{\mu}=D/d\tau=\dot{\left(\,\,\right)}$,
\begin{equation}
\theta=\nabla_{\mu}u^{\mu}\,,\label{eq:expasnion}
\end{equation}
is the expansion (note that it is consistent with \eqref{eq:teta})
and 
\begin{equation}
\bot_{\mu\nu}=g_{\mu\nu}-u_{\mu}u_{\nu}\,,\label{eq:projector}
\end{equation}
 is the projector to the hypersurface orthogonal to $u^{\mu}$. 

Clearly our fluid is irrotational - the twist tensor is vanishing
so that for the extrinsic curvature we get 
\begin{equation}
K_{\mu\nu}=\perp_{\mu}^{\lambda}\nabla_{\lambda}u_{\nu}=\perp_{\nu}^{\lambda}\nabla_{\lambda}u_{\mu}=K_{\nu\mu}\,.\label{eq:no_Twist}
\end{equation}
Moreover, because of the constraint \eqref{eq:constaraint} the four
velocity $u^{\mu}$ is tangential to the time-like geodesics \cite{Lim:2010yk}
\begin{equation}
a^{\mu}=\dot{u}^{\mu}=u^{\lambda}\nabla_{\lambda}u^{\mu}=\nabla^{\lambda}\phi\nabla_{\lambda}\nabla^{\mu}\phi=\frac{1}{2}\nabla^{\mu}\left(\nabla^{\lambda}\phi\nabla_{\lambda}\phi\right)=0\,.
\end{equation}
Therefore we will call the $\partial_{\mu}\varphi$ frame \textendash{}
the natural or the geodesic frame. However, from \ref{eq:Noether decomposed}
it follows that the shift-charges do not move along the geodesics.
If $\gamma$ is a constant parameter the current \eqref{eq:Noether decomposed}
is conserved 
\[
\nabla_{\mu}J^{\mu}=0\,,
\]
 we obtain 
\[
\gamma\Box\theta-2\dot{\lambda}-2\lambda\theta=0\,,
\]
which is nothing else as the equation of motion. This equation we
can rewrite as 
\begin{equation}
\gamma\left(\ddot{\theta}-\nabla^{2}\theta\right)-2\dot{\lambda}-2\lambda\theta=0\,,\label{eq:Klein_Gordon for the expansion}
\end{equation}
where 
\begin{equation}
\nabla^{2}=-\perp_{\mu}^{\lambda}\nabla_{\lambda}\left(\perp_{\alpha}^{\mu}\nabla^{\alpha}\right)=-\vec{\nabla}_{\mu}\vec{\nabla}^{\mu}\,,\label{eq:Laplacian}
\end{equation}
 denotes the spatial Laplacian. Decomposing the EMT \eqref{eq:EMT_General}
in the same way we obtain 
\begin{equation}
T_{\mu\nu}=\varepsilon u_{\mu}u_{\nu}-p\perp_{\mu\nu}+q_{\mu}u_{\nu}+q_{\nu}u_{\mu}\,,\label{eq:EMT fluid}
\end{equation}
where the energy density is 
\begin{equation}
\varepsilon=T_{\mu\nu}u^{\mu}u^{\nu}=2\lambda-\gamma\left(\dot{\theta}-\frac{1}{2}\theta^{2}\right)\,,\label{eq:energy density}
\end{equation}
the pressure 
\begin{equation}
p=-\frac{1}{3}T^{\mu\nu}\perp_{\mu\nu}=-\gamma\left(\dot{\theta}+\frac{1}{2}\theta^{2}\right)\,,\label{eq:Pressure}
\end{equation}
and the energy flux 
\begin{equation}
q_{\mu}=\bot_{\mu\lambda}T_{\sigma}^{\lambda}u^{\sigma}=-\gamma\bot_{\mu}^{\lambda}\nabla_{\lambda}\theta=\bot_{\mu}^{\lambda}J_{\lambda}\,.\label{eq:energy flux}
\end{equation}
Thus in the LRF the energy is transported along the spatial gradients
of the expansion. However, there is no heat flux because $q_{\mu}=\perp_{\mu\nu}J^{\nu}$.
Similarly to the imperfect fluid from \emph{Kinetic Gravity Braiding}
\cite{Pujolas:2011he} we can use this fact to claim that the Landau-Lifshitz
frame moving with energy coincides up to $\mathcal{O}\left(\gamma\right)$
with the Eckart frame moving with the shift-charges, see \cite{Israel:1976tn},
Eq. (2), p. 312. One can also see that the anisotropic stress is vanishing
in this frame 
\begin{equation}
\Pi_{\mu\nu}=\left(\perp_{\mu\alpha}\perp_{\nu\beta}-\frac{1}{3}\perp_{\mu\nu}\perp_{\alpha\beta}\right)T^{\alpha\beta}=0\,.\label{eq:general_anisotropic}
\end{equation}

Further one can observe that an analog of the Euler relation holds:
\begin{equation}
\varepsilon+p=n-\gamma\dot{\theta}\,,\label{eq:Euler}
\end{equation}
where the last terms on the r.h.s. appears purely because of the imperfect
character of the fluid. Note that this expression corresponds to the
unit chemical potential, because of the normalisation of our field.
From the expressions \eqref{eq:energy density}, \eqref{eq:Pressure}
and \eqref{eq:Euler} it follows that for nearly neutral $n\simeq0$
system with a slow expansion a $\dot{\theta}\ll\theta^{2}$ the equation
of state is approximately that of vacuum $p\simeq-\varepsilon$. This
happens even, if the mimetic fluid is not a dominant source of curvature. 

Further we will need the Raychaudhuri equation for the timelike geodesics
\begin{equation}
\dot{\theta}=-\frac{1}{3}\theta^{2}-\sigma_{\mu\nu}\sigma^{\mu\nu}-R_{\mu\nu}u^{\mu}u^{\nu}\,,\label{eq:Raychaudhuri}
\end{equation}
where $\sigma_{\mu\nu}$ is the shear tensor
\[
\sigma_{\mu\nu}=\frac{1}{2}\left(\bot_{\mu}^{\lambda}\nabla_{\lambda}u_{\nu}+\bot_{\nu}^{\lambda}\nabla_{\lambda}u_{\mu}\right)-\frac{1}{3}\bot_{\mu\nu}\theta\,,
\]
$\sigma^{2}=\sigma_{\mu\nu}\sigma^{\mu\nu}$. Thus we see that in
the presence of the external matter the energy density and pressure
of the mimetic fluid depend on the pressure and energy density of
this external matter through the Einstein equations in the form 
\[
R_{\mu\nu}=T_{\mu\nu}-\frac{1}{2}Tg_{\mu\nu}\,.
\]
These equations give us 
\begin{equation}
R_{\mu\nu}u^{\mu}u^{\nu}=\frac{\varepsilon+3p}{2}+\frac{\rho_{\text{ext}}+3P_{\text{ext}}}{2}=\lambda-2\gamma\dot{\theta}-\frac{\gamma}{2}\theta^{2}+\frac{\rho_{\text{ext}}+3P_{\text{ext}}}{2}\,,\label{eq:R_uu}
\end{equation}
where for the external matter $\rho_{\text{ext}}$ and $P_{\text{ext}}$
are defined in the LRF associated with $u^{\mu}$. Thus using the
Raychaudhuri equation \eqref{eq:Raychaudhuri} we can express the
time derivative of the expansion 
\[
\dot{\theta}\left(1-2\gamma\right)=-\lambda-\frac{1}{3}\left(1-\frac{3\gamma}{2}\right)\theta^{2}-\sigma^{2}-\frac{\rho_{\text{ext}}+3P_{\text{ext}}}{2}\,.
\]
Now we can write the corresponding formulas for the energy density
\eqref{eq:energy density}: 
\[
\varepsilon=\left(\frac{2-3\gamma}{1-2\gamma}\right)\lambda+\frac{\gamma}{6}\left(\frac{5-9\gamma}{1-2\gamma}\right)\theta^{2}+\frac{\gamma}{1-2\gamma}\left(\sigma^{2}+\frac{\rho_{\text{ext}}+3P_{\text{ext}}}{2}\right)\,,
\]
and pressure \eqref{eq:Pressure}: 
\[
p=\frac{\gamma}{1-2\gamma}\lambda-\frac{\gamma}{6}\left(\frac{1-3\gamma}{1-2\gamma}\right)\theta^{2}+\frac{\gamma}{1-2\gamma}\left(\sigma^{2}+\frac{\rho_{\text{ext}}+3P_{\text{ext}}}{2}\right)\,.
\]

It is important to note that the local energy density and pressure
of DM explicitly depend the external matter e.g. on baryonic energy
density. This could potentially explain dependance of the observable
halo profiles on the local baryonic physics, (especially if we consider
$\gamma\left(\varphi\right)$), see the corresponding discussion on
this issue in \cite{Khoury:2014tka}. We can also differentiate the
Raychaudhuri equation to obtain express $\dot{\lambda}$. For our
time-like geodesics the Raychaudhuri equation for the shear tensor
after some calculations takes the form 
\[
\dot{\sigma}_{\mu\nu}=-\frac{1}{3}\bot_{\mu\nu}\dot{\theta}-K_{\mu}^{\alpha}K_{\alpha\nu}+\frac{1}{2}u^{\alpha}u^{\beta}\bot_{\mu}^{\lambda}\bot_{\nu}^{\gamma}\left(R_{\beta\gamma\lambda\alpha}+R_{\beta\lambda\gamma\alpha}\right)\,,
\]
where we have used the notation \eqref{eq:no_Twist}. Differentiating
the Raychaudhuri for the expansion we obtain 
\[
-\dot{\lambda}=\left(1-2\gamma\right)\ddot{\theta}+\left(\frac{2}{3}-\gamma\right)\theta\dot{\theta}+2\dot{\sigma}_{\mu\nu}\sigma^{\mu\nu}+\frac{\dot{\rho}_{\text{ext}}+3\dot{P}_{\text{ext}}}{2}\,,
\]
which we can plug into the equation of motion \eqref{eq:Klein_Gordon for the expansion}
to obtain for the highest derivatives of this equation 
\begin{equation}
\ddot{\theta}\left(2-3\gamma\right)-\gamma\nabla^{2}\theta+2u^{\alpha}u^{\beta}\sigma^{\mu\nu}\, C_{\alpha\mu\nu\beta}+...=0\,,\label{eq:Non_Linear_Wave}
\end{equation}
where the ellipsis stands for the terms with the lower number of derivatives
and $C_{\alpha\mu\nu\beta}$ is the Weyl tensor and $\nabla^{2}$
is the spatial Laplacian \eqref{eq:Laplacian}. From this equation
one can see that the speed of propagation of the expansion - sound
speed is 
\[
c_{\text{S}}^{2}=\frac{\gamma}{2-3\gamma}\,,
\]
exactly as it was calculated for the cosmological case in \cite{Chamseddine:2014vna},
provided that the shear and the Weyl tensors are vanishing. The coupling
between the shear and the Weyl tensor hints to the possible change
of the speed of propagation for the gravitons on the backgrounds with
non-vanishing shear. However, caution is needed, as the Einstein equations
do not contain higher derivatives of the expansion, therefore similarly
to \cite{Deffayet:2010qz} the system is triangular and there is no
change in the speed of propagation for the gravitons. Moreover, because
of the anisotropy given by $q_{\mu}$ one can expect that on general
backgrounds with shear and Weyl tensor the phonons of the mimetic
fluid propagate with different speeds in different directions even
in the LRF. This issues definitely require further investigation using
the original dynamical variable $\delta\varphi$. Especially this
anisotropy in the sound speed can be important for studies of the
nonlinear collapse.

\subsection{Other frames }

As we have shown above the EMT of the mimetic fluid does not take
the form of a perfect fluid. Instead, it has a form of a fluid with
an energy flow. The natural frame $u_{\mu}=\partial_{\mu}\varphi$
is particular, because i) in this frame there is no anisotropic stress,
ii) there is no vorticity, iii) this frame moves along timelike geodesics,
iv) the relation of the energy density, pressure etc to the field
$\varphi$ and its derivatives is polynomial. Thus in this geodesic
frame all equations look less nonlinear than in other frames. It is
important to stress that for \emph{any non-ideal fluid (EMT)} one
can use the geodesic frame considered above. However, neither shift-charges
nor energy move along this velocity field. One should expect that
the velocity fields corresponding to the motion of charges and energy
are more important in a physical setup e.g. for a consideration of
structure formation or nonlinear gravitational collapse.

\subsubsection{Eckart frame }

Another useful frame is the so-called Eckart frame. This is the frame
moving together with the charges (provided the current \eqref{eq:Noether decomposed}
is timelike and future-directed). In this frame we have 
\begin{equation}
U_{\mu}=\frac{J_{\mu}}{n_{\text{E}}}=\frac{nu_{\mu}+q_{\mu}}{n_{\text{E}}}=\frac{2\lambda\partial_{\mu}\varphi-\gamma\partial_{\mu}\theta}{n_{\text{E}}}\simeq\partial_{\mu}\varphi-\frac{\gamma}{2\lambda}\bot_{\mu}^{\lambda}\partial_{\lambda}\theta\,,\label{eq:Eckart_Frame}
\end{equation}
where $n_{\text{E}}$ is the proper density of shift-charges
\begin{equation}
n_{\text{E}}=\sqrt{J^{\alpha}J_{\alpha}}=\sqrt{n^{2}-q^{2}}=\sqrt{4\lambda^{2}-4\lambda\gamma\dot{\theta}+\gamma^{2}\left(\partial\theta\right)^{2}}\simeq2\lambda\left(1-\frac{\gamma}{2\lambda}\dot{\theta}\right)\,,\label{eq:Eckart_Charge}
\end{equation}
 where 
\begin{equation}
q^{2}=-q^{\mu}q_{\mu}\,.\label{eq:q_mod}
\end{equation}
Further it is convenient to introduce a unit spacelike vector in the
direction of the energy transfer 
\[
\hat{q}_{\mu}=q_{\mu}/q\,.
\]
In this frame the EMT is 
\begin{equation}
T_{\mu\nu}=\left(\mathcal{E}_{\text{E}}+P_{\text{E}}\right)U_{\mu}U_{\nu}-P_{\text{E}}g{}_{\mu\nu}+\Pi_{\mu\nu}^{\text{E}}+Q_{\mu}U_{\nu}+Q_{\nu}U_{\mu}\,,\label{eq:EMT_Eckart}
\end{equation}
where the notation is the same as in the formulas \eqref{eq:energy density},
\eqref{eq:Pressure}, \eqref{eq:energy flux}, \eqref{eq:projector}.
The four-velocity of charges $U^{\mu}$ is obtained \eqref{eq:Eckart_Frame}
from $u^{\mu}$ by the Lorentz transformation 
\[
U_{\mu}=\frac{nu_{\mu}+q\hat{q}_{\mu}}{n_{\text{E}}}=\frac{u_{\mu}+v\hat{q}_{\mu}}{\sqrt{1-v^{2}}}\,,
\]
where the relative velocity is given by 
\begin{equation}
v=\frac{q}{n}\,,\label{eq:Eckart_relative_velocity}
\end{equation}
so that as expected 
\begin{equation}
n=\frac{n_{\text{\text{E}}}}{\sqrt{1-v^{2}}}\,.\label{eq:Eckart_Charge_density}
\end{equation}
Thus the transition to the Eckart frame is the Lorentz boost $\left(u_{\mu},\hat{q}_{\mu}\right)\rightarrow\left(U_{\mu},\hat{Q}_{\mu}\right)$
with \foreignlanguage{english}{ 
\begin{eqnarray}
u_{\mu}=\frac{U_{\mu}-v\hat{Q}_{\mu}}{\sqrt{1-v^{2}}}\,, & \mbox{and} & \hat{q}_{\mu}=\frac{\hat{Q}_{\mu}-vU_{\mu}}{\sqrt{1-v^{2}}}\,,\label{eq:uq}
\end{eqnarray}
}and\foreignlanguage{english}{ 
\begin{equation}
\hat{Q}_{\mu}=\frac{vu_{\mu}+\hat{q}_{\mu}}{\sqrt{1-v^{2}}}\,.\label{eq:Qunit}
\end{equation}
}By plugging in the expressions \eqref{eq:uq} into the EMT \eqref{eq:EMT fluid}
we obtain EMT from \eqref{eq:EMT_Eckart} with 
\begin{equation}
\Pi_{\mu\nu}^{\text{E}}=\beta\left(\hat{Q}_{\mu}\hat{Q}_{\nu}+\frac{1}{3}\left(g_{\mu\nu}-U_{\mu}U_{\nu}\right)\right)\,,\label{eq:Anisotropic_Eckart}
\end{equation}
\begin{equation}
Q_{\mu}=-\left(q+\frac{\beta}{v}\right)\hat{Q}_{\mu}\,,\label{eq:Q_Eckart}
\end{equation}
\begin{equation}
\mathcal{E}_{\text{E}}=\varepsilon+\beta\,,\label{eq:E_Eckart}
\end{equation}
\begin{equation}
P_{\text{E}}=p+\frac{1}{3}\beta\,,\label{eq:P_Eckart}
\end{equation}
where we denoted 
\begin{equation}
\beta=v^{2}\left(\frac{\varepsilon+p-2n}{1-v^{2}}\right)=-2\lambda\left(\frac{v^{2}}{1-v^{2}}\right)\,.\label{eq:Betta}
\end{equation}
 In the leading order in $\gamma$ we obtain: 
\begin{equation}
\beta\simeq-2\lambda v^{2}=-\frac{2\lambda q^{2}}{n^{2}}\simeq\frac{\gamma^{2}\perp^{\mu\nu}\partial{}_{\mu}\theta\,\partial_{\nu}\theta}{2\lambda}\,,\label{eq:Betta_leading}
\end{equation}
so that %
\footnote{This implies that $\beta<0$ for most natural physical cases when
the leading energy density is positive $\lambda>0$. %
}, 
\begin{equation}
Q_{\mu}\simeq\frac{\gamma\dot{\theta}}{n}q_{\mu}\simeq-\gamma^{2}\frac{\dot{\theta}\vec{\nabla}_{\mu}\theta}{2\lambda}\,,\label{eq:Q_Eckart_Leading}
\end{equation}
and 
\begin{equation}
\Pi_{\mu\nu}^{\text{E}}\simeq\beta\left(\hat{q}_{\mu}\hat{q}_{\nu}+\frac{1}{3}\perp_{\mu\nu}\right)\,.\label{eq:Anisotrop_Eckart_leading}
\end{equation}
It is important that energy transfer and anisotropic stress are both
$\mathcal{O}\left(\gamma^{2}\right)$ and quadratic in $\theta$.
Now we can compare the anisotropic stress with the usual shear viscosity.
In the leading order 
\[
\sigma_{\mu\nu}^{\text{E}}\simeq\sigma_{\mu\nu}=\phi_{;\mu;\nu}-\frac{1}{3}\theta\perp_{\mu\nu}\,,
\]
so that 
\[
\Pi_{\mu\nu}^{\text{E}}\simeq-\frac{\beta}{\theta}\left(\sigma_{\mu\nu}-\pi_{\mu\nu}\right)\,,
\]
with an additional part 
\[
\pi_{\mu\nu}=\theta\hat{q}_{\mu}\hat{q}_{\nu}+\phi_{;\mu;\nu}\,.
\]
Both tensors $\sigma_{\mu\nu}$ and $\pi_{\mu\nu}$ are symmetric,
traceless and purely spatial. Because these tensors are not positive-definite
generically they cannot be simultaneously diagonalized. However, if
$\left\Vert \pi_{\mu\nu}\right\Vert \ll\left\Vert \sigma_{\mu\nu}\right\Vert $
the anisotropic stress mimics effects of shear viscosity with the
shear viscosity coefficient 
\[
\eta\simeq\frac{\beta}{\theta}=\frac{\gamma^{2}}{2\lambda}\left(\frac{\perp^{\mu\nu}\partial_{\mu}\theta\partial_{\nu}\theta}{\theta}\right)\,.
\]
In cosmology we can estimate these terms as 
\[
Q_{\mu}\sim-\gamma^{2}\frac{\dot{H}\vec{\nabla}_{\mu}\zeta}{H^{2}}\,,
\]
 where $\zeta$ is the curvature perturbation. Hence the magnitude
of the energy transfer on physical scale $\ell$ is suppressed as
\[
Q_{\ell}\sim c_{\text{S}}^{4}\frac{\dot{H}\xi}{H}\left(\ell H\right)^{-1}\,.
\]
The anisotropic stress only comes at quadratic order in perturbations.
In particular, 
\[
\eta\sim-\frac{\gamma^{2}}{H}\,\delta_{ik}\left(\frac{\partial_{i}\xi}{aH}\right)\left(\frac{\partial_{i}\xi}{aH}\right)\,,
\]
and on physical scale $\ell$ its magnitude is strongly suppressed
\[
\eta_{\ell}\sim-\frac{c_{\text{S}}^{4}}{H}\xi_{\ell}^{2}\left(\ell H\right)^{-2}\,.
\]

\subsubsection{Landau-Lifshitz frame }

Let us find the so-called Landau-Lifshitz (LL) frame $V^{\mu}$ in
which there is no energy flow - so that $V^{\mu}$ is a timelike,
unit and future directed eigenvector of the EMT - $T_{\nu}^{\mu}V^{\mu}=\mathcal{E}V^{\nu}$
where $\mathcal{E}$ is the energy density in this frame. The results
one can find e.g. in \cite{Sawicki:2012pz} where the same structure
of the EMT was considered. One obtains for the eigenvalue 
\[
\mathcal{E}=\frac{\varepsilon-p}{2}+\sqrt{\left(\frac{\varepsilon+p}{2}\right)-q^{2}}\,,
\]
 and the corresponding 4-velocity is 
\[
V_{\mu}\propto u_{\mu}+v\hat{q}_{\mu}=\varphi_{,\mu}\left(1+\frac{\gamma v}{q}\dot{\theta}\right)-\frac{\gamma v}{q}\theta_{,\mu}\,,
\]
 and the relative velocity between the LL frame and the natural geodesic
frame is 
\begin{equation}
v_{\text{LL}}=\frac{\varepsilon+p}{2q}-\sqrt{\left(\frac{\varepsilon+p}{2q}\right)^{2}-1}\,.\label{eq:relative_V_LL}
\end{equation}
In this frame the EMT is 
\[
T_{\mu\nu}=\mathcal{E}V_{\mu}V_{\nu}-P_{\text{LL}}\bot_{\mu\nu}^{\text{LL}}+\Pi_{\mu\nu}^{\text{LL}}\,,
\]
where the anisotropic stress $\Pi_{\mu\nu}^{\text{LL}}=\mathcal{O}\left(\gamma^{2}\right)$.
Note that the difference in the relative velocities of the Eckart
and Landau-Lifshitz frames is $v-v_{\text{LL}}=\mathcal{O}\left(\gamma^{2}\right)$.

\subsection{Gradient expansion }

The mimetic fluid shares some similarities with the imperfect fluid
from \emph{Kinetic Gravity Braiding} \cite{Pujolas:2011he}. In particular,
there is a mixing with the gravity and the energy flux forcing the
EMT to deviate from the perfect fluid form. Similarly \cite{Pujolas:2011he}
it is useful to look at the (spatial) gradient expansion around the
equilibrium also for the mimetic fluid. For one can make a boost to
the new LRF -the Eckart frame \eqref{eq:Eckart_Frame} moving together
with the charges: 
\[
U_{\mu}\simeq\frac{J_{\mu}}{n}\simeq u_{\mu}+\frac{q_{\mu}}{n}\,,
\]
where we have suppressed terms $\mathcal{O}\left(\gamma^{2}\right)$
as we will further constantly do further in this section. It is important
to stress that the Eckart frame does have vorticity, because $\partial_{\mu}\phi$
and $n^{-1}\gamma\bot_{\mu}^{\lambda}\nabla_{\lambda}\theta$ are
not proportional to a common gradient, see \eqref{sub:Vorticity}.In
this frame the EMT takes the form of the perfect fluid 
\[
T_{\mu\nu}\simeq\left(\varepsilon+p\right)U_{\mu}U_{\nu}-pg_{\mu\nu}+\mathcal{O}\left(\gamma^{2}\right)\,,
\]
where $p$ and $\varepsilon$ are still defined as in \eqref{eq:energy density}
and \eqref{eq:Pressure}. From equations \eqref{eq:R_uu} and \eqref{eq:Raychaudhuri}
we obtain that
\[
\dot{\theta}\simeq-\frac{1}{3}\theta^{2}-\sigma_{\mu\nu}\sigma^{\mu\nu}-\lambda-\frac{\rho_{\text{ext}}+3P_{\text{ext}}}{2}\,,
\]
 
\[
p\simeq\gamma\lambda=\frac{\gamma}{2}\varepsilon\,,
\]
 thus up to $\mathcal{O}\left(\gamma^{2}\right)$ terms $c_{\text{S}}^{2}=\gamma/2$.
Here we neglected terms with expansion $\theta$, because they are
higher order in the gradient expansion $\mathcal{O}\left(\partial^{2}\right)$.
Also we neglected the external matter contribution, because this is
a purely gravitational effect. For the charge density we get 
\[
n\simeq\left(2+\gamma\right)\lambda\,,
\]
and for the energy density we get 
\[
\varepsilon\simeq2\lambda\,.
\]
Using the formalism from \cite{Schutz} we can write the chemical
potential $\mu$ as 
\[
U_{\mu}=\mu^{-1}\left(\partial_{\mu}\phi-n^{-1}\gamma\bot_{\mu}^{\lambda}\nabla_{\lambda}\theta\right)\simeq\mu^{-1}\left(\partial_{\mu}\phi-\frac{\gamma}{2\lambda}\partial_{\mu}\theta\right)\,,
\]
so that $V^{\mu}V_{\mu}=1$ and 
\[
\mu\simeq1+\frac{\gamma}{2\lambda}\dot{\theta}\simeq1-\frac{\gamma}{2}=\frac{\partial\varepsilon}{\partial n}\,,
\]
where in the last equality we have neglected higher order derivatives
and the external contribution $\rho_{\text{ext}}+3P_{\text{ext}}$.
Thus we have demonstrated that \emph{mimetic fluid} with HD in the
gradient expansion corresponds to the perfect fluid with equation
of state $\gamma/2$. In this approximation, any deviation from the
perfect fluid appears only on the level $\mathcal{O}\left(\gamma^{2}\right)$
or through Planck-scale suppressed operators.

\subsection{Vorticity\label{sub:Vorticity} }

As we have seen both Landau-Lifshitz and Eckart frames have the form
\begin{equation}
U_{\mu}=\mu^{-1}\left(\partial_{\mu}\varphi+\rho\partial_{\mu}\theta\right)\,,\label{eq:Clebsch}
\end{equation}
were $\mu$ and $\rho$ are some scalar functions which together with
$\theta$ and $\varphi$ build relativistic Clebsch potentials, see
e.g. \cite{Schutz}. Thus the vorticity vector $\Omega^{\mu}\left(V\right)$
for any timelike vector field $V^{\mu}$ is defined as 
\[
\Omega^{\mu}\left(V\right)=\frac{1}{2}\varepsilon^{\alpha\beta\gamma\mu}V_{\gamma}\bot_{\alpha}^{\lambda}V_{\beta;\lambda}=\frac{1}{2}\varepsilon^{\alpha\beta\gamma\mu}V_{\gamma}V_{\beta;\alpha}\,,
\]
where $\bot_{\mu\nu}=g_{\mu\nu}-V_{\mu}V_{\nu}$. For the velocity
field \eqref{eq:Clebsch} we have 
\[
\Omega^{\mu}\left(U\right)=\frac{1}{2\mu^{2}}\varepsilon^{\alpha\beta\gamma\mu}\rho_{,\alpha}\,\theta_{,\beta}\,\varphi_{,\gamma}=\frac{1}{2\mu^{2}}\varepsilon^{\alpha\beta\gamma\mu}\vec{\nabla}_{\alpha}\rho\,\vec{\nabla}_{\beta}\theta\, u_{\gamma}\,,
\]
where $\vec{\nabla}_{\alpha}=\bot_{\alpha}^{\lambda}\nabla_{\lambda}$
and $\bot_{\mu\nu}=g_{\mu\nu}-u_{\mu}u_{\nu}$ where we have used
the geodesic frame. 

In particular, in the Eckart frame we have $\rho=-\gamma/2\lambda$
and $\mu=n_{\text{E}}/2\lambda$ so that 
\[
\Omega_{\text{E}}^{\mu}=\frac{\gamma}{n_{\text{E}}^{2}}\varepsilon^{\alpha\beta\gamma\mu}\vec{\nabla}_{\alpha}\lambda\,\vec{\nabla}_{\beta}\theta\, u_{\gamma}\simeq\frac{\gamma}{4\lambda^{2}}\varepsilon^{\alpha\beta\gamma\mu}\vec{\nabla}_{\alpha}\lambda\,\vec{\nabla}_{\beta}\theta\, u_{\gamma}\,,
\]
where in the last equality we omitted terms of the order $\mathcal{O}\left(\gamma^{2}\right)$.
Generically the vorticity does not vanish, because $\lambda$ and
$\theta$ correspond to different initial data and can be chosen independently.
Indeed, without HD, $\lambda\left(\mathbf{x}\right)$ corresponds
to the initial energy density profile, while $\theta\left(\mathbf{x}\right)$
is fixed by the initial velocities and the metric. As we have already
mentioned the Eckart frame and Landau-Lifshitz frame coincide up to
factors $\mathcal{O}\left(\gamma\right)$. Thus the vorticity is also
there in the Landau-Lifshitz frame.

\section{Conclusions and Open Questions\label{sec:Conclusions-and-Open}}

On the level of cosmological background we found that \emph{mimetic
fluid} with higher derivatives (HD), $\gamma\left(\Box\varphi\right)^{2}$,
renormalises the Newton\textquoteright{}s constant in the Friedmann
equations by $\delta G_{\text{N}}/G_{\text{N}}\sim\gamma$. We have
discussed different bounds on this renormalisation coming from observations,
see subsection \ref{sub:Bounds-on-the}. These bounds are not particularly
severe. Further we have proved that \emph{mimetic fluid} has zero
pressure when there is no extra matter in the universe. Surprisingly
our \emph{imperfect} DM survives the inflationary expansion and builds
a small part of the dominating energy density e.g. radiation after
that. However, inflation completely redshifts the shift-charge so
that after that the shift-symmetric system is on the dynamical attractor
solution perfectly tracking \emph{any} external matter. Thus, if there
is no external matter, the mimetic fluid on attractor with zero shift-charge
does not have any energy density and cannot be DM. If one insists
on the mimetic origin \cite{Chamseddine:2013kea} of the constraint
$\varphi_{,\mu}\varphi^{,\mu}=1$ and on shift-symmetry, then the
system cannot be DM. We have showed that a short period of shift-symmetry
breaking in the HD term (through $\gamma\left(\varphi\right)$) during
the radiation domination époque removes this problem. During this
cosmologically short period of time one can create sufficient amount
of shift-charges to enter the usual matter domination époque at the
usual redshift. The charge creation should happen in that case at
temperatures $T_{\text{cr}}\gtrsim0.1\,\mbox{MeV}$ for the current
DM sound speed $c_{\text{S}}^{2}\lesssim10^{-5}$ and at $T_{\text{cr}}\gtrsim10\,\mbox{GeV}$
for the $c_{\text{S}}^{2}\simeq10^{-10}$ chosen in \cite{Capela:2014xta}
to suppress perturbations with wavelengths below 100 kpc. It is important
to note that this model essentially has three parameters \textendash{}
the sound speed in the early universe (or $\gamma_{\text{early}}$),
the sound speed in the late universe (or $\gamma_{\text{late}}$)
and the duration of the shift-symmetry breaking $\Delta t$ which
is cosmologically short and would be rather difficult to observe. 

We have also demonstrated that the shift-symmetry breaking in form
of potential $V\left(\varphi\right)$ would not help to generate enough
shift-charge. On the other hand one can use the phase of shift-symmetry
breaking to generate the cosmological constant $\Lambda$. It would
be very interesting to see whether this DM picture can be elegantly
combined with inflation or with Dark Energy similarly to \cite{Lim:2010yk}. 

We have shown that the HD operator $\gamma\left(\varphi\right)\left(\Box\varphi\right)^{2}$
modifies the simplest irrotational DM to an \emph{imperfect fluid}
with a finite sound speed $c_{\text{S}}^{2}$ of the order $\mathcal{O}\left(\gamma\right)$,
finite vorticity of the order $\mathcal{O}\left(\gamma\right)$ in
the frames moving either with the shift-charges or with the energy.
In the natural rest frame of the field $u_{\mu}=\partial_{\mu}\varphi$
the fluid has the energy flow $\mathcal{O}\left(\gamma\right)$. Generically
there is no reference frame where the fluid would look locally isotropic.
Thus one can expect that around a general configuration the sound
speed depends on direction in all reference frames. This can change
the whole structure formation picture. On the other hand we presented
an argument that for all backgrounds with vanishing shear and Weyl
tensor the speed of sound is the same as in cosmology. If the shear
is not vanishing there is an intriguing possibility that the speed
of propagation of gravity waves changes around such configurations
and becomes anisotropic, see \eqref{eq:Non_Linear_Wave}. 

Moreover, the appearance of the vorticity, energy flow and anisotropic
stress should make the collapse by far less efficient as it is in
the case for the irrotational DM \cite{Sawicki:2013wja}, \cite{Izumi:2009ry}.
Thus it seems that this scalar field theory of DM is not bound to
be a small part of the total DM budget. There is still an open issue
with caustics which are quite often formed in the hydrodynamical and
in particular scalar models of DM, see e.g. \cite{Felder:2002sv,ArkaniHamed:2005gu,Blas:2009yd,Mukohyama:2009tp}.
Indeed, it is not clear whether this imperfect DM forms caustics,
other nonlinear singularities or supports strong shock waves. Clearly
\emph{mimetic} DM without higher derivatives will form caustics more
efficient than standard rotational DM. The HD (or dissipative) terms
have potential to avoid the caustics in some cases \cite{ArkaniHamed:2005gu}.
Moreover, on galactic scales DM is virialized and this implies many
caustics on the fluid level. Thus it is not clear how to model such
strong overdensities in a fluid-like picture and not to form caustics.
A proper interpretation of a caustic in the field-theoretical setup
is needed. It may happen that the only way to continue the solution
through the multivalued region is the UV physics or quantisation of
the system. Another related problem is to find the strong coupling
scale for the quantised sound waves around a general configuration.
Further it is important to understand the Hamiltonian formulation
of this theory. The first step without HD was done in \cite{Chaichian:2014qba}.
It is known that HD can substantially change the structure of the
theory also in the presence of constraints, see e.g. \cite{Chen:2012au}.
These issues lie beyond the scope of this paper but definitely requires
a further investigation. 

Depending on the value of the sound speed, the linear perturbation
theory of the mimetic imperfect DM can have rather interesting phenomenological
consequences related to the finiteness of the sound speed and the
corresponding suppression of the power spectrum on the scales shorter
than the sonic horizon, see \cite{Capela:2014xta}.

We think that the \emph{imperfect} DM provides a perfect playground
for a phenomenologically rich and sophisticated modeling of the universe
where we live in.

\section*{Acknowledgements\label{sec:Acknowledgements}}

\addcontentsline{toc}{section}{Acknowledgements} 

It is a pleasure to thank Elias Kiritsis, Viatcheslav Mukhanov, Sabir
Ramazanov, Ignacy Sawicki, Sergey Sibiryakov and Peter Tinyakov for
very useful discussions and criticisms. We are also indebted to Sabir
Ramazanov and Fabio Capela for sharing the preliminary results of
their research presented in \cite{Capela:2014xta}. AV is thankful
to the theory group of the Université Libre de Bruxelles for a very
warm hospitality during the final stages of this project.

\bibliographystyle{utcaps}
\addcontentsline{toc}{section}{\refname}\bibliography{Mimetic}

\end{document}